# Anisotropic Multi-layer Cylindrical Structures Containing Graphene Layers: An Analytical Approach

**Mohammad Bagher Heydari [1,*], Mohammad Hashem Vadjed Samiei [1]**

[1,*] School of Electrical Engineering, Iran University of Science and Technology (IUST), Tehran, Iran

[*]Corresponding author: mo_heydari@alumni.iust.ac.ir ; heydari.sharif@gmail.com

**Abstract** We propose a novel analytical model for anisotropic multi-layer cylindrical structures containing graphene layers. The general structure is formed by an aperiodic repetition of a three-layer sub-structure, where a graphene layer, with isotropic surface conductivity of σ, has been sandwiched between two adjacent magnetic materials. Each anisotropic material has the permittivity and permeability tensors of $\bar{\bar{\varepsilon}}$ and $\bar{\bar{\mu}}$, respectively. An external magnetic bias has been applied in the axial direction. General matrix representation is obtained in our proposed analytical model to find the dispersion relation. The relation will be used to find the effective index of the structure and its other propagation parameters. Two special exemplary structures have been introduced and studied to show the richness of the proposed general structure regarding the related specific plasmonic wave phenomena and effects. A series of simulations have been conducted to demonstrate the noticeable wave-guiding properties of the structure in the 10-40 THz band. A very good agreement between the analytical and simulation results is observed. The proposed structure can be utilized to design novel plasmonic devices such as absorbers, modulators, plasmonic sensors and tunable antennas in the THz frequencies.

## 1. Introduction

Graphene plasmonics is a new emerging science, which studies the excitation of Surface Plasmon Polaritons (SPPs) on the graphene sheets and their applications for designing THz devices such as such as waveguides [1-13], isolator [14], circulator [15, 16], coupler [17], resonator [18], antennas [19-21], filter [22], Radar Cross-Section (RCS) reduction-based devices [23-25], and graphene-based medical components [26-32]. It should be noted that noble metals support SPPs at the near-infrared and visible frequencies [18, 33-42].
. This science is developed based on the optical conductivity of the graphene, which allows one to control the plasmonic features of the device via electrostatic or magnetostatic gating. Among the fascinating plasmonic devices, the cylindrical plasmonic structures have been attracted the attention of researchers, due to their interesting applications such as absorbers [43], cloaking [44], Faraday rotation [45], fiber [46], modulator [47], sensor [48] and reconfigurable antennas [49], in the THz region.

Graphene-based Cylindrical Waveguides (GCWs) have been addressed in some articles [50-61]. In [53], the dispersion curves for the hybrid modes of graphene-coated silicon nano-wire have been depicted and discussed, where the authors have considered the permittivity of the silicon by the second-order polynomial. The complex waves have been classified as trapped surface waves, fast and slow leaky waves and their characteristics are studied more precisely [53]. Jian-Ping Liu and his co-workers have studied surface plasmons in a hybrid type of GCW in [56, 57]. In [56], their proposed waveguide had only one graphene layer, while two graphene layers have been applied in the studied structure of [57]. The plasmonic guiding properties of hybrid GCW have been discussed in [56, 57], where it has been shown that the waveguides have some fascinating advantages such as long propagation length and low mode area. One of the novel articles in this debated field has been presented by Dmitry A. Kuzmin et al. [61], which discusses the propagation of TE modes in GCWs at visible frequencies. The authors have calculated the critical value for the radius of the waveguide, which is required for supporting TE-plasmons. In [60], the whispering gallery modes of graphene-based InGaAs nanowire have been considered and a high quality-factor of 235 for a 5 nm radius has been reported.

To author's knowledge, a comprehensive study on anisotropic multilayered cylindrical structures with graphene layers has not been reported in any published article. This paper aims to present a general graphene-based cylindrical structure and its analytical model to cover all special cases of graphene-based cylindrical waveguides. Our proposed structure is composed of a graphene layer sandwiched between two magnetic materials. Each material has the permittivity and permeability tensors of $\bar{\bar{\varepsilon}}$ and $\bar{\bar{\mu}}$, respectively. The external magnetic bias is applied in the z-direction. Since the direction of the applied bias is parallel to the graphene layers, thus they have isotropic surface conductivities. Our general and complex structure allows someone to design tunable and controllable plasmonic components, which are adjustable by changing the magnetic bias and the chemical potential. It should be noted that anisotropic materials or anisotropic metamaterials have many fascinating applications in the literature [62-66]. For instance, a left-handed material slab waveguide was designed and studied in [62], which had a negative group velocity. In [64, 66], a slab waveguide has been reported for sensor applications by utilizing an anisotropic metamaterial.

The paper is organized as follows. Section 2 presents a novel analytical model for our proposed general structure. The dispersion matrix for the general structure will be derived in this section, which can be used to obtain plasmonic features such as the effective index. To verify the analytical model outlined in section 2 and also show the richness of the proposed structure, two exemplary structures are studied in section 3. The first waveguide is a graphene-coated nano-wire, where the graphene is deposited on the $SiO_2$ nano-wire. The second example is a hybridization of graphene and the gyro-electric substrate. This hybridization leads to study tunable non-reciprocal plasmonic features in this structure, which we believe that it will be helpful for designing new devices in THz frequencies. Finally, section 4 concludes the article.

## 2. The Proposed General Structure and the Formulation of the Problem

This section proposes a novel analytical model for cylindrically anisotropic multi-layered structures containing graphene layers. Firstly, we will find the dispersion relation of the general structure. Then, obtaining plasmonic features of the structure, such as the effective index and propagation loss is straightforward. Fig. 1 represents the schematic of the proposed structure, where the graphene layer has been sandwiched between two adjacent magnetic materials, each one has the permittivity and permeability tensors of $\bar{\bar{\varepsilon}}$ and $\bar{\bar{\mu}}$, respectively. The electric and magnetic currents have been located at the outer cylindrical surface of the structure. The structure has been magnetized along the z-direction by a DC bias magnetic field $B_0$. Since the external magnetic field has been applied parallel to the graphene layer, its conductivity in the N-th layer has the familiar relation of Kubo's formula [67]:

$$\sigma_N(\omega,\mu_{c,N},\Gamma_N,T) = \frac{-je^2}{4\pi\hbar}Ln\left[\frac{2|\mu_{c,N}|-(\omega-j2\Gamma_N)\hbar}{2|\mu_{c,N}|+(\omega-j2\Gamma_N)\hbar}\right] + \frac{-je^2 K_B T}{\pi\hbar^2(\omega-j2\Gamma_N)}\left[\frac{\mu_{c,N}}{K_B T} + 2Ln\left(1+e^{-\mu_{c,N}/K_B T}\right)\right] \quad (1)$$

Where $\hbar$ is the reduced Planck's constant, $K_B$ is Boltzmann's constant, ω is radian frequency, $e$ is the electron charge, $\Gamma_N$ is the phenomenological electron scattering rate for that layer ($\Gamma_N = 1/\tau_N$, where $\tau_N$ is the relaxation time), T is the temperature, and $\mu_{c,N}$ is the chemical potential for the N-th layer which can be altered by the chemical doping or electrostatic bias [67]. It should be mentioned that the graphene layer can be modeled as a material with the following permittivity relation:

$$\varepsilon_{g,N} = 2.5 + j\frac{\sigma_N}{\omega\varepsilon_0 \Delta_N} \quad (2)$$

Where 2.5 is the surface-normal effective permittivity of the graphene and $\Delta_N$ is the thickness of the N-th graphene layer. The permeability and permittivity tensors of the N-th layer of bi-anisotropic medium, in the presence of the DC magnetic bias in the z-direction, are expressed as the following tensor [68]:

$$\bar{\bar{\mu}}_N = \mu_0 \begin{pmatrix} \mu_N & j\mu_{a,N} & 0 \\ -j\mu_{a,N} & \mu_N & 0 \\ 0 & 0 & \mu_{\|,N} \end{pmatrix} \tag{3}$$

$$\bar{\bar{\varepsilon}}_N = \varepsilon_0 \begin{pmatrix} \varepsilon_N & j\varepsilon_{a,N} & 0 \\ -j\varepsilon_{a,N} & \varepsilon_N & 0 \\ 0 & 0 & \varepsilon_{\|,N} \end{pmatrix} \tag{4}$$

Where $\varepsilon_0$ and $\mu_0$ are the permittivity and permeability of the free space, respectively.

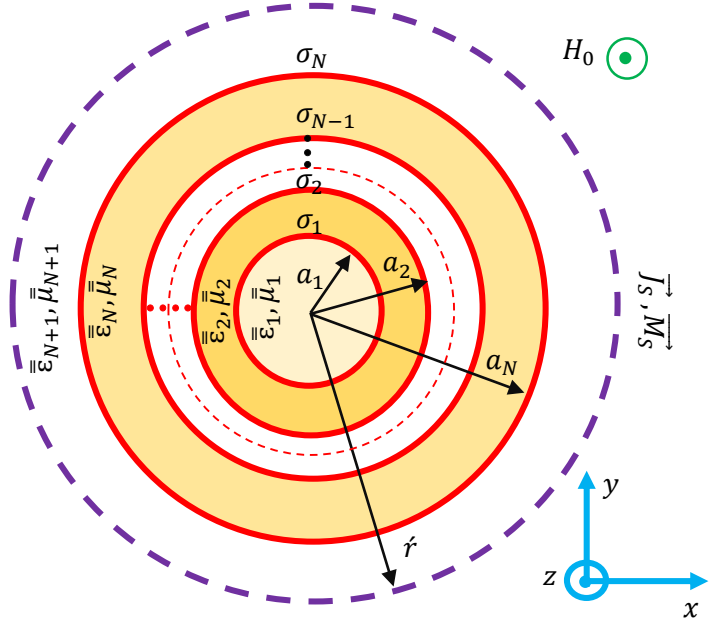

**Fig. 1.** The cross-section of a general anisotropic multi-layer cylindrical structure containing electrically biased graphene layers. The electric and magnetic currents have been located at the outer cylindrical surface of the structure.

It should be noted that the diagonal and off-diagonal elements of the permeability tensor for the magnetic materials have usual forms, written here for the *N*-th layer [68]:

$$\mu_N = 1 + \frac{\omega_{M,N}(\omega_{H,N} + j\omega\alpha_N)}{\omega_{H,N}^2 - (1+\alpha_N^2)\omega^2 + 2j\alpha_N\omega\omega_{H,N}} \tag{5}$$

$$\mu_{a,N} = \frac{\omega\omega_{M,N}}{\omega_{H,N}^2 - (1+\alpha_N^2)\omega^2 + 2j\alpha_N\omega\omega_{H,N}} \tag{6}$$

$$\mu_{\|,N} = 1 - \frac{j\alpha_N\omega_{M,N}}{\omega + j\alpha_N\omega_{H,N}} \tag{7}$$

In the above relations, $\gamma_N$ is the gyromagnetic ratio and $\alpha_N$ is the Gilbert damping constant for the *N*-th layer. Moreover, $\omega_{H,N} = \gamma_N H_0$ and $\omega_M = \gamma_N M_S$, $M_S$ is the saturation magnetization. In the literature, the diagonal and off-diagonal elements of permittivity tensors for the bi-gyrotropic media are considered as the following tensor, where have been expressed for the *N*-th layer [69]:

$$\varepsilon_N = \varepsilon_{\infty,N}\left(1 - \frac{\omega_{p,N}^2(\omega + j\upsilon_N)}{\omega\left[(\omega + j\upsilon_N)^2 - \omega_{c,N}^2\right]}\right) \tag{8}$$

$$\varepsilon_{a,N} = \varepsilon_{\infty,N}\left(\frac{\omega_{p,N}^2 \omega_c}{\omega\left[(\omega + j\upsilon_N)^2 - \omega_c^2\right]}\right) \tag{9}$$

$$\varepsilon_{\parallel,N} = \varepsilon_{\infty,N}\left(1 - \frac{\omega_{p,N}^2}{\omega(\omega + j\upsilon_N)}\right) \tag{10}$$

In (8)-(10), $\upsilon_N$ is the effective collision rate and $\varepsilon_{\infty,N}$ is the background permittivity of the $N$-th layer. Furthermore, the plasma and the cyclotron frequencies are defined as follows [69]:

$$\omega_{p,N} = \sqrt{\frac{n_s e^2}{\varepsilon_0 \varepsilon_{\infty,N} m^*}} \tag{11}$$

$$\omega_c = \frac{e B_0}{m^*} \tag{12}$$

Where $e$, $m^*$ and $n_s$ are the charge, effective mass and the density of the carriers. Now, consider Maxwell's equations inside the bi-gyrotropic media of the $N$-th layer in the frequency domain (suppose $e^{i\omega t}$) [68]:

$$\nabla \times \mathbf{E} = -j\omega \bar{\bar{\mu}}_N \cdot \mathbf{H} \tag{13}$$

$$\nabla \times \mathbf{H} = j\omega \bar{\bar{\varepsilon}}_N \cdot \mathbf{E} \tag{14}$$

By utilizing the above equations in the cylindrical coordinates, the z-component of the electric and magnetic fields inside the bi-gyrotropic layer satisfy [68]:

$$\left(\nabla_\perp^2 + \frac{\varepsilon_\parallel}{\varepsilon}\frac{\partial^2}{\partial z^2} + (k_0^2 \varepsilon_\parallel \mu_\perp)\right)E_z + k_0 \mu_\parallel \left(\frac{\varepsilon_a}{\varepsilon} + \frac{\mu_\alpha}{\mu}\right)\frac{\partial}{\partial z} H_z = 0 \tag{15}$$

$$\left(\nabla_\perp^2 + \frac{\mu_\parallel}{\mu}\frac{\partial^2}{\partial z^2} + (k_0^2 \varepsilon_\perp \mu_\parallel)\right)H_z - k_0 \varepsilon_\parallel \left(\frac{\varepsilon_a}{\varepsilon} + \frac{\mu_\alpha}{\mu}\right)\frac{\partial}{\partial z} E_z = 0 \tag{16}$$

where

$$\nabla_\perp^2 = \frac{1}{r}\frac{\partial}{\partial r} r \frac{\partial}{\partial r} + \frac{1}{r^2}\frac{\partial}{\partial^2 \varphi} \tag{17}$$

In these equations, $k_0$ is the free space wave-number and,

$$\varepsilon_{\perp,N} = \varepsilon_N - \frac{\varepsilon_{a,N}^2}{\varepsilon_N} \tag{18}$$

$$\mu_{\perp,N} = \mu_N - \frac{\mu_{a,N}^2}{\mu_N} \tag{19}$$

Consider the plasmonic waves propagating inside the bi-gyrotropic media in the z-direction, which the z-component of the electric and magnetic fields are expressed as follows:

$$H_{z,N}(r,\varphi,z) = \int_{-\infty}^{+\infty} \sum_{m=-\infty}^{\infty} e^{jk_z z} \exp(-jm\varphi) H_{z,N}^m(\text{r}) dk_z \tag{20}$$

$$E_{z,N}(r,\varphi,z) = \int_{-\infty}^{+\infty} \sum_{m=-\infty}^{\infty} e^{jk_z z} \exp(-jm\varphi) E_{z,N}^m(\mathrm{r}) dk_z \tag{21}$$

In the above equations, $m$ is an integer and $k_z$ is the propagation constant. Now, by substituting (20) and (21) into (15) and (16), one can obtain the following coupled equations:

$$\left( \nabla_\perp^2 + k_0^2 \varepsilon_{\|,N} \mu_{\perp,N} - \frac{\varepsilon_{\|,N}}{\varepsilon_N} k_z^2 \right) E_{z,N}^m - jk_0 k_z \mu_{\|,N} \left( \frac{\varepsilon_{a,N}}{\varepsilon_N} + \frac{\mu_{\alpha,N}}{\mu_N} \right) H_{z,N}^m = 0 \tag{22}$$

$$\left( \nabla_\perp^2 + k_0^2 \varepsilon_{\perp,N} \mu_{\|,N} - \frac{\mu_{\|,N}}{\mu_N} k_z^2 \right) H_{z,N}^m + jk_0 k_z \varepsilon_{\|,N} \left( \frac{\varepsilon_{a,N}}{\varepsilon_N} + \frac{\mu_{\alpha,N}}{\mu_N} \right) E_{z,N}^m = 0 \tag{23}$$

Then, a fourth-order differential equation is achieved by combining (22) and (23),

$$\left( \nabla_\perp^2 + k_{r,2N-1}^2 \right)\left( \nabla_\perp^2 + k_{r,2N}^2 \right) H_{z,N}^m = 0 \tag{24}$$

By considering the following coefficients,

$$A_{1,N} = -k_0^2 \left( \varepsilon_{\|,N} \mu_{\perp,N} + \varepsilon_{\perp,N} \mu_{\|,N} \right) + \left( \frac{\varepsilon_{\|,N}}{\varepsilon_N} + \frac{\mu_{\|,N}}{\mu_N} \right) k_z^2 \tag{25}$$

$$A_{2,N} = \left( k_0^2 \varepsilon_{\|,N} \mu_{\perp,N} - \frac{\varepsilon_{\|,N}}{\varepsilon_N} k_z^2 \right)\left( k_0^2 \varepsilon_{\perp,N} \mu_{\|,N} - \frac{\mu_{\|,N}}{\mu_N} k_z^2 \right) - k_0^2 k_z^2 \mu_{\|,N} \varepsilon_{\|,N} \left( \frac{\varepsilon_{a,N}}{\varepsilon_N} + \frac{\mu_{\alpha,N}}{\mu_N} \right)^2 \tag{26}$$

The characteristics equation of (24) is written for the $N$-th layer

$$s^4 + A_{1,N} s^2 + A_{2,N} = 0 \tag{27}$$

Next, the roots of characteristics equation for each medium ($N$-th layer) are derived

$$k_{r,2N-1} = \sqrt{\frac{-A_{1,N} + \sqrt{A_{1,N}^2 - 4A_{2,N}}}{2}} \tag{28}$$

$$k_{r,2N} = \sqrt{\frac{-A_{1,N} - \sqrt{A_{1,N}^2 - 4A_{2,N}}}{2}} \tag{29}$$

Therefore, the roots of characteristics equations for various regions of Fig.1 are considered as follows:

$$k_r = \begin{cases} k_{r,1}, k_{r,2} & i = 1,2 \; ; \; \text{for layer } N = 1 \\ \ldots & \ldots \\ k_{r,2N-1}, k_{r,2N} & i = 2N-1, 2N \; ; \; \text{for layer } N \\ k_{r,2N+1}, k_{r,2N+2} & i = 2N+1, 2N+2 \; ; \; \text{for layer } N+1 \end{cases} \tag{30}$$

In (30), $N$ denotes the number of the layer and $i$ indicates the index of the roots for that layer. Now, one should write the electromagnetic fields $H_m(z)$ and $E_m(z)$ in various regions,

$$H_z^m(\mathrm{r}) = \begin{cases} A_{m,1,1} J_m(k_{r,1}r) + B_{m,2,1} J_m(k_{r,2}r) & r < a_1 \\ A_{m,3,2} J_m(k_{r,3}r) + B_{m,3,2} H_m^{(2)}(k_{r,3}r) + \\ A_{m,4,2} J_m(k_{r,4}r) + B_{m,4,2} H_m^{(2)}(k_{r,4}r) & a_1 < r < a_2 \\ \ldots \\ A_{m,2N+1,N+1} J_m(k_{r,2N+1}r) + B_{m,2N+1,N+1} H_m^{(2)}(k_{r,2N+1}r) + \\ A_{m,2N+2,N+1} J_m(k_{r,2N+2}r) + B_{m,2N+2,N+1} H_m^{(2)}(k_{r,2N+2}r) & a_N < r < r' \\ C_{m,2N+1,N+1} H_m^{(2)}(k_{r,2N+1}r) + D_{m,2N+2,N+1} H_m^{(2)}(k_{r,2N+2}r) & r > r' \end{cases} \quad (31)$$

$$E_z^m(\mathrm{r}) = \begin{cases} A_{m,1,1} T_{1,1} J_m(k_{r,1}r) + B_{m,2,1} T_{2,1} J_m(k_{r,2}r) & r < a_1 \\ T_{3,2}\left(A_{m,3,2} J_m(k_{r,3}r) + B_{m,3,2} H_m^{(2)}(k_{r,3}r)\right) + \\ T_{4,2}\left(A_{m,4,2} J_m(k_{r,4}r) + B_{m,4,2} H_m^{(2)}(k_{r,4}r)\right) & a_1 < r < a_2 \\ \ldots \\ T_{2N+1,N+1}\left(A_{m,2N+1,N+1} J_m(k_{r,2N+1}r) + B_{m,2N+1,N+1} H_m^{(2)}(k_{r,2N+1}r)\right) + \\ T_{2N+2,N+1}\left(A_{m,2N+2,N+1} J_m(k_{r,2N+2}r) + B_{m,2N+2,N+1} H_m^{(2)}(k_{r,2N+2}r)\right) & a_N < r < r' \\ T_{2N+1,N+1} C_{m,2N+1,N+1} H_m^{(2)}(k_{r,2N+1}r) + T_{2N+2,N+1} D_{m,2N+2,N+1} H_m^{(2)}(k_{r,2N+2}r) & r > r' \end{cases} \quad (32)$$

Where

$$T_{i,N} = \frac{1}{-jk_0 k_z \varepsilon_{\|,N}\left(\dfrac{\varepsilon_{a,N}}{\varepsilon_N} + \dfrac{\mu_{\alpha,N}}{\mu_N}\right)}\left(k_{r,i}^2 - k_0^2 \varepsilon_{\perp,N}\mu_{\|,N} + \frac{\mu_{\|,N}}{\mu_N}k_z^2\right) \quad i = 2N-1, 2N \; ; N = 1, 2, \ldots \quad (33)$$

are the coefficients that are appeared in (32). The transverse components of electric and magnetic fields are expressed as:

$$\begin{pmatrix} E_{r,N} \\ H_{r,N} \end{pmatrix} = \bar{\bar{Q}}_N^{Pos} \frac{\partial}{\partial r}\begin{pmatrix} E_{z,N} \\ H_{z,N} \end{pmatrix} + \frac{m}{r}\bar{\bar{Q}}_N^{Neg}\begin{pmatrix} E_{z,N} \\ H_{z,N} \end{pmatrix} \quad (34)$$

$$j\begin{pmatrix} E_{\varphi,N} \\ H_{\varphi,N} \end{pmatrix} = \bar{\bar{Q}}_N^{Neg} \frac{\partial}{\partial r}\begin{pmatrix} E_{z,N} \\ H_{z,N} \end{pmatrix} + \frac{m}{r}\bar{\bar{Q}}_N^{Pos}\begin{pmatrix} E_{z,N} \\ H_{z,N} \end{pmatrix} \quad (35)$$

Where the Q-matrices in (34) and (35) have been defined as following matrices,

$$\bar{\bar{Q}}_N^{Pos} = \frac{1}{2}\left[\frac{1}{-k_z^2 + k_0^2 \varepsilon_{+,N}\mu_{+,N}}\begin{pmatrix} jk_z & -\omega\mu_0\mu_{+,N} \\ \omega\varepsilon_0\varepsilon_{+,N} & jk_z \end{pmatrix} + \frac{1}{-k_z^2 + k_0^2 \varepsilon_{-,N}\mu_{-,N}}\begin{pmatrix} jk_z & \omega\mu_0\mu_{-,N} \\ -\omega\varepsilon_0\varepsilon_{-,N} & jk_z \end{pmatrix}\right] \quad (36)$$

$$\bar{\bar{Q}}_N^{Neg} = \frac{1}{2}\left[\frac{1}{-k_z^2 + k_0^2\varepsilon_{+,N}\mu_{+,N}}\begin{pmatrix} jk_z & -\omega\mu_0\mu_{+,N} \\ \omega\varepsilon_0\varepsilon_{+,N} & jk_z \end{pmatrix} - \frac{1}{-k_z^2 + k_0^2\varepsilon_{-,N}\mu_{-,N}}\begin{pmatrix} jk_z & \omega\mu_0\mu_{-,N} \\ -\omega\varepsilon_0\varepsilon_{-,N} & jk_z \end{pmatrix}\right]$$
(37)

Moreover,

$$\varepsilon_{\pm,N} = \varepsilon_N \pm \varepsilon_{a,N} \tag{38}$$

$$\mu_{\pm,N} = \mu_N \pm \mu_{a,N} \tag{39}$$

Now, let us apply the boundary conditions to obtain the characteristics equation of the proposed structure. For the graphene layer sandwiched between two magnetic materials, the boundary conditions are written in general form:

$$E_{z,N} = E_{z,N+1}, E_{\varphi,N} = E_{\varphi,N+1} \qquad N = 1,2,3,.... \tag{40}$$

$$H_{z,N+1} - H_{z,N} = -\sigma E_{\varphi,N}, H_{\varphi,N+1} - H_{\varphi,N} = \sigma E_{z,N} \qquad N = 1,2,3,.... \tag{41}$$

And for the last boundary at $r = \acute{r}$,

$$E_{z,N+1}^{>} - E_{z,N+1}^{<} = M_{s\varphi}, E_{\varphi,N+1}^{>} - E_{\varphi,N+1}^{<} = -M_{sz} \tag{42}$$

$$H_{z,N+1}^{>} - H_{z,N+1}^{<} = -J_{s\varphi}, H_{\varphi,N+1}^{>} - H_{\varphi,N+1}^{<} = J_{sz} \tag{43}$$

In (42) and (43), $M_{sz}, M_{s\varphi}, J_{sz}, J_{s\varphi}$ are $z$ and $\varphi$-components of magnetic and electric currents at $r = \acute{r}$, respectively. By applying the boundary conditions expressed in (40)-(43), the final matrix representation for our general waveguide is obtained,

$$\bar{\bar{S}}_{4N+4,4N+4} \cdot \begin{pmatrix} A_{m,1,1} \\ B_{m,2,1} \\ A_{m,3,2} \\ B_{m,3,2} \\ ... \\ A_{m,2N+2,N+1} \\ B_{m,2N+2,N+1} \\ C_{m,2N+2,N+1} \\ D_{m,2N+2,N+1} \end{pmatrix}_{4N+4,1} = \begin{pmatrix} 0 \\ 0 \\ 0 \\ 0 \\ ... \\ M_{s\varphi} \\ -M_{sz} \\ -J_{s\varphi} \\ J_{sz} \end{pmatrix}_{4N+4,1} \tag{44}$$

In (44), the matrix $\bar{\bar{S}}$ is:

$$\bar{\bar{S}}_{4N+4,4N+4} = \begin{pmatrix} T_{1,1}J_m(k_{r,1}a_1) & ... & ... & -T_{3,2}H_m^{(2)}(k_{r,3}a_1) & ... & 0 & 0 & 0 & 0 \\ R_{1,1,1}^{J}(a_1) & ... & ... & -R_{3,2,1}^{H}(a_1) & ... & 0 & 0 & 0 & 0 \\ ... & ... & ... & ... & ... & 0 & 0 & 0 & 0 \\ ... & ... & ... & ... & ... & 0 & 0 & 0 & 0 \\ ... & ... & ... & ... & ... & ... & ... & ... & ... \\ 0 & 0 & 0 & 0 & ... & ... & ... & ... & ... \\ 0 & 0 & 0 & 0 & ... & ... & ... & ... & ... \\ 0 & 0 & 0 & 0 & ... & ... & ... & ... & ... \\ 0 & 0 & 0 & 0 & ... & ... & ... & ... & -R_{2N+2,N+2,2}^{H}(\text{r}') \end{pmatrix} \tag{45}$$

Where the following relations have been used in (45),

$$\begin{pmatrix} R^{J}_{i,N,1}(\mathrm{r}) \\ R^{J}_{i,N,2}(\mathrm{r}) \end{pmatrix} = -j\left[ \bar{\bar{Q}}^{Neg}_{N} k_{r,i} J'_{m}(k_{r,i}r) + \frac{m}{r} \bar{\bar{Q}}^{Pos}_{N} J_{m}(k_{r,i}r) \right] \begin{pmatrix} T_{i,N} \\ 1 \end{pmatrix} \quad i = 2N-1, 2N; N = 1,2,\ldots \quad (46)$$

$$\begin{pmatrix} R^{H}_{i,N,1}(\mathrm{r}) \\ R^{H}_{i,N,2}(\mathrm{r}) \end{pmatrix} = -j\left[ \bar{\bar{Q}}^{Neg}_{N} k_{r,i} H'^{(2)}_{m}(k_{r,i}r) + \frac{m}{r} \bar{\bar{Q}}^{Pos}_{N} H^{(2)}_{m}(k_{r,i}r) \right] \begin{pmatrix} T_{i,N} \\ 1 \end{pmatrix} \quad i = 2N-1, 2N; N \geq 2 \quad (47)$$

Now, our analytical model has been completed for the general structure of Fig. 1. It should be mentioned that the matrix $\bar{\bar{S}}$ is an important matrix, since it obtains the dispersion relation or the propagation constant of the structure by setting $det(\bar{\bar{S}}) = 0$. In the next step, the plasmonic parameters of the general multi-layer structure such as the effective index ($n_{eff} = Re(k_z/k_0)$), the propagation length ($L_{Prop} = 1/2Im(k_z)$), and figure of merit based on the quality factor (or briefly called "benefit-to-cost ratio", $FOM = Re(k_z)/2\pi Im(k_z)$) [70] is straightforward. In what follows, we will consider two exemplary structural variants to show the richness of the proposed general structure regarding the related specific plasmonic wave phenomena and effects.

## 3. Results and Discussion

In this section, two graphene-based cylindrical waveguides, as special cases of the general proposed structure, have been studied to show, first the validity and accuracy of the proposed model, and second the richness of the proposed general waveguide regarding the related specific plasmonic wave phenomena and effects. The first waveguide is a well-known structure, a graphene-coated nano-wire, to check the validity and the performance of the proposed analytical model. As the second example, a new anisotropic multi-layer cylindrical waveguide is introduced and studied, constituting of Graphene-InSb-SiO$_2$-Si layers. The anisotropic layer is n-type InSb, with anisotropic permittivity tensor. This waveguide supports tunable SPPs, their plasmonic properties are altered by the magnetic bias and the chemical potential. In this section, the first two modes ($m = 0,1$) are investigated to be brief. Furthermore, the graphene thickness is $\Delta = 0.5\ nm$, the temperature is supposed to be *T=300 K*, and the relaxation time of the graphene is $\tau = 0.45\ ps$ in all simulations.

*3.1 The First Structure: The Graphene-coated Nano-wire*

Fig. 2 illustrates the schematic of the first example, where the graphene is deposited on the SiO$_2$ nano-wire with a permittivity of $\varepsilon_{SiO_2} = 2.09$ and the radius of $R_{SiO_2}$. Without loss of generality and for simplicity, we assume that the surrounding medium is air ($\varepsilon_0, \mu_0$). Here, we set the nano-wire radius $R_{SiO_2} = 90\ nm$ and the chemical potential $\mu_c = 0.45\ eV$ unless otherwise stated. In this paper, we have neglected the effect of the optical phonon on the plasmonic features in our studied frequency range. Hence, the graphene conductivity is modeled by using relation (2).

The dispersion relation of the nano-wire is obtained by utilizing the matrix representation of (45). To show the validity of the proposed model outlined in the previous section, the plasmonic properties of the nano-waveguide, including the effective index defined as $n_{eff} = Re(k_z/k_0)$ and the propagation length ($L_{Prop} = 1/2Im(k_z)$), have been illustrated for the first two modes ($m = 0,1$) in Fig. 3. The exponential factor of $e^{-jm\varphi}$ in relation (20) and (21) represents the order of the plasmonic modes. It is worth to mention that the propagating modes inside the nano-waveguide are only TM modes since TE modes cannot be excited in the chemical potential range of 0.1~1 *eV* in our studied frequency region. An excellent agreement is seen between the theoretical and simulation results in Fig. 3, which indicates the high accuracy of the proposed analytical model. It is obvious that the mode $m = 0$ is cutoff-free and the plasmonic mode $m = 1$ is excited only for $n_{eff} < \sqrt{\varepsilon_{SiO_2}}$. As the frequency increases, the effective index increases, because the plasmonic wave is concentrated on the graphene layer. However, the propagation loss increases for high frequencies results in the reduction of the propagation length. As a result, there is a trade-off between the effective index and the propagation length at the specific frequency.

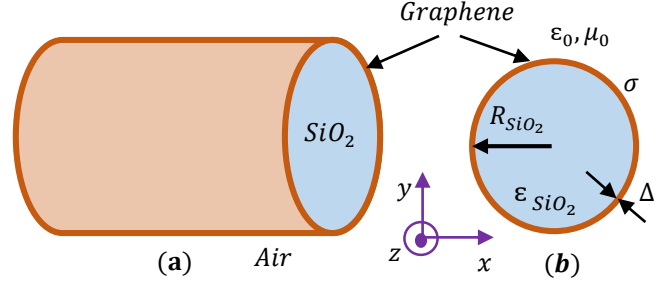

**Fig. 2.** (a) The 3D structure, and (b) the cross-section view of the graphene-based nano-waveguide.

The Figure of Merit, defined as $FOM = Re(k_z)/2\pi Im(k_z)$ [70], is one of the powerful parameters for investigating the performance of the plasmonic structures. In Fig. 4, the FOM curves as a function of the chemical potential and the nano-wire radius have been depicted for the first two modes at the frequency of 25 THz. For the cutoff-free mode ($m = 0$), it is obvious from Fig. 4(a) that better FOM is achievable for high values of the chemical doping. There is an optimum value of FOM at the specific chemical potential for the mode $m = 1$, amounting to 20 e.g., at about 0.65 eV. One can observe that the fundamental mode propagates inside the plasmonic nano-waveguide for each value of the cylinder radius, which means that this mode has not cut-off radius. To design a nano-wire operating as single-mode, the radius of the waveguide must be $R_{SiO_2} < 75\ nm$. For instance, the structure works as single-mode for $R_{SiO_2} = 60\ nm$ with $FOM = 35$ at the frequency of 25 THz. Fig. 4(b) indicates that the mode $m = 1$ has better FOM than the cutoff-free mode for high values of the nanowire radius ($R_{SiO_2} > 170\ nm$).

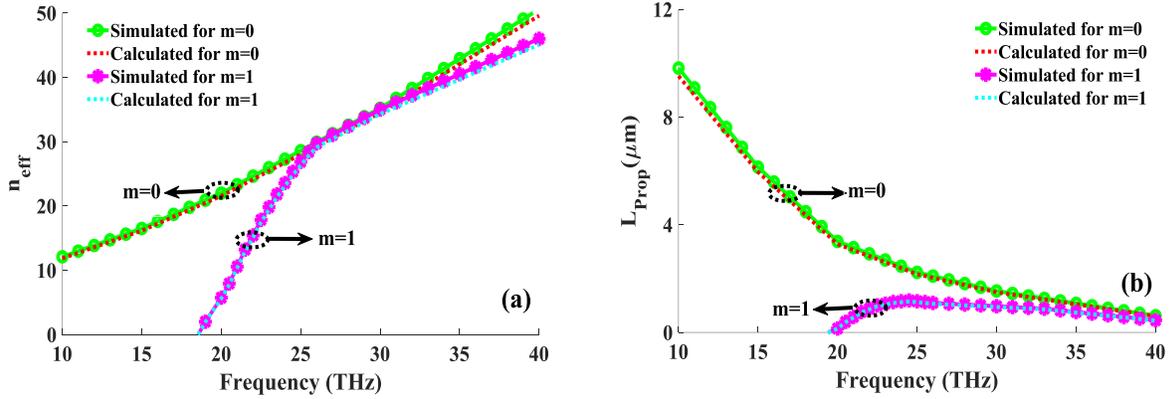

**Fig. 3.** The analytical and simulation results for the modal properties of the nano-waveguide: **(a)** the effective index, **(b)** the propagation length. The diagrams have been depicted for the first two modes ($m = 0,1$).

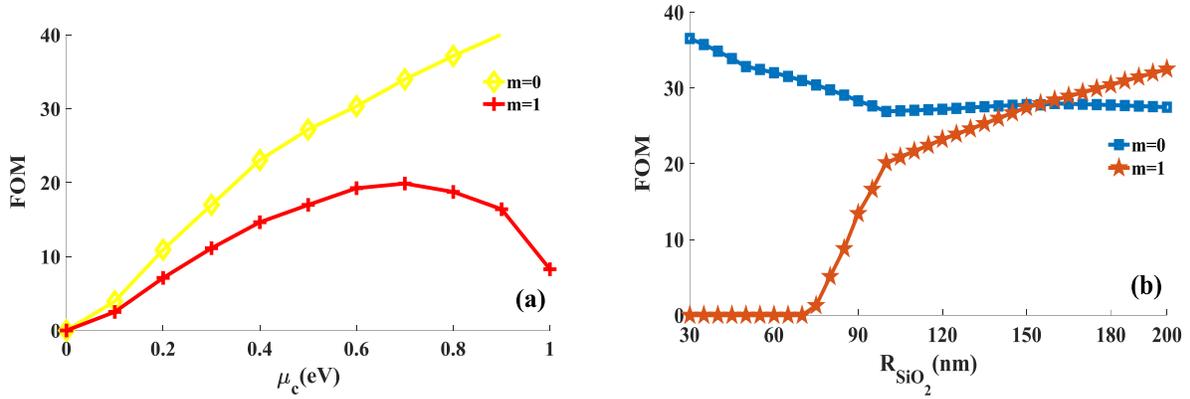

**Fig. 4.** The analytical results of FOM as a function of: **(a)** the chemical potential of the graphene, **(b)** the radius of the nano-wire. The operation frequency is supposed to be 25 THz here. The radius of the structure is 90 nm in diagram (a) and the chemical potential is assumed to be 0.45 eV in (b).

*3.2 The Second Structure: The Graphene-Based Cylindrical Waveguide with Gyro-electric Substrate*

As the second example, a novel graphene-based structure with the gyro-electric substrate is introduced and studied, as shown in Fig. 5. In this structure, the magnetic bias is applied in the z-direction. To simulate the proposed structure, the gyroelectric substrate is assumed to be n-type InSb with the thickness of $t_{InSb} = R_{InSb} - R_{SiO_2} = 5\ nm$, which its parameters are $\mu_2 = \mu_0, \varepsilon_\infty = 15.68,\ m^* = 0.022 m_e, n_s = 1.07 \times 10^{17}/cm^3, \nu = 0.314 \times 10^{13} s^{-1}$ and $m_e$ is the electron's mass. Without loss of generality and for simplicity, we presume that the surrounding medium is air and the gyroelectric substrate is located on SiO$_2$-Si layers ( $\varepsilon_{Si} = 11.9,\ \varepsilon_{SiO_2} = 2.09$). The geometrical parameters are supposed $R_{Si} = 30\ nm, t_{SiO_2} = R_{SiO_2} - R_{Si} = 3\ nm$ and the chemical potential of the graphene is $\mu_c = 0.7\ eV$ unless otherwise stated.

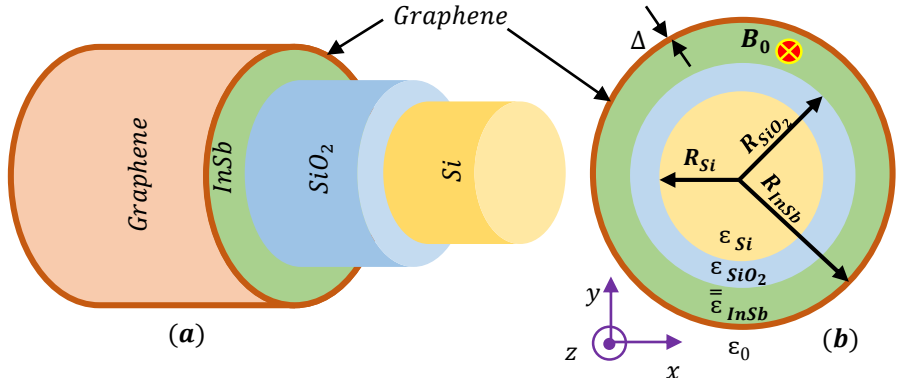

**Fig. 5.** The schematic of the graphene-based cylindrical waveguide with an anisotropic substrate: **(a)** the 3D structure, **(b)** the cross-section view of the structure. The external magnetic field is applied in the z-direction.

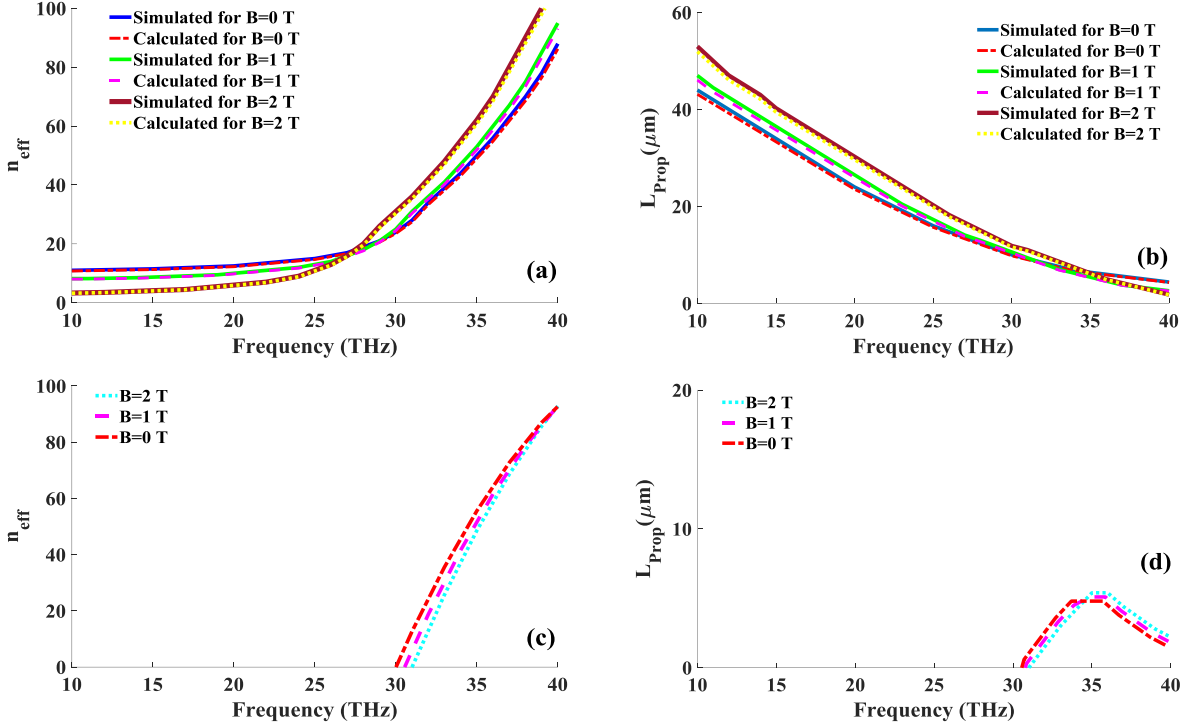

**Fig. 6. (a), (b)** The analytical and simulation results for the effective index and the propagation length of the fundamental mode ($m = 0$) for various magnetic bias ($B_0 = 0,1,2$), **(c), (d)** The analytical results for the effective index and the propagation length of the first mode ($m = 1$) for various magnetic bias ($B_0 = 0,1,2$). The chemical potential is 0.7 eV ($\mu_c = 0.7\ eV$).

The plasmonic parameters such as the effective index and the propagation length for the waveguide are derived by using the proposed analytical model outlined in the previous section (see relation (45)). Here, we do not mention and report these equations due to their complicated mathematical forms. The modal properties of the fundamental mode ($m = 0$) and the 1'st mode ($m = 1$) for various external magnetic fields are illustrated in Fig. 6. To better representation, they have been depicted for $m = 0$ and $m = 1$ in different diagrams. We should mention that the propagating modes inside the structure are hybrid TM-TE modes in general, due to the usage of the gyroelectric layer.

There is a full agreement between the theoretical and simulation results in Fig. 6, which confirms the validity of the proposed analytical model. It is evident that the fundamental mode ($m = 0$) is a cut-off free mode, while the 1'st mode ($m = 1$) has a cut-off frequency, varies by changing the magnetic bias. For instance, the cut-off frequency occurs at 32 THz for the external bias of 2 T. One can observe that the effective index increases, as the frequency increases. But the propagation length has an opposite trend. It should be emphasized that one of the main properties of the proposed structure is its ability to tune the modal properties via the magnetic bias. As seen in Fig. 6(a), the increment of the magnetic bias has a great influence on the effective index of the cut-off free mode for $f > 35\ THz$. As observed in Fig. 6(d), the mode $m = 1$ has low propagation length, which cannot propagate for large nano-distances.

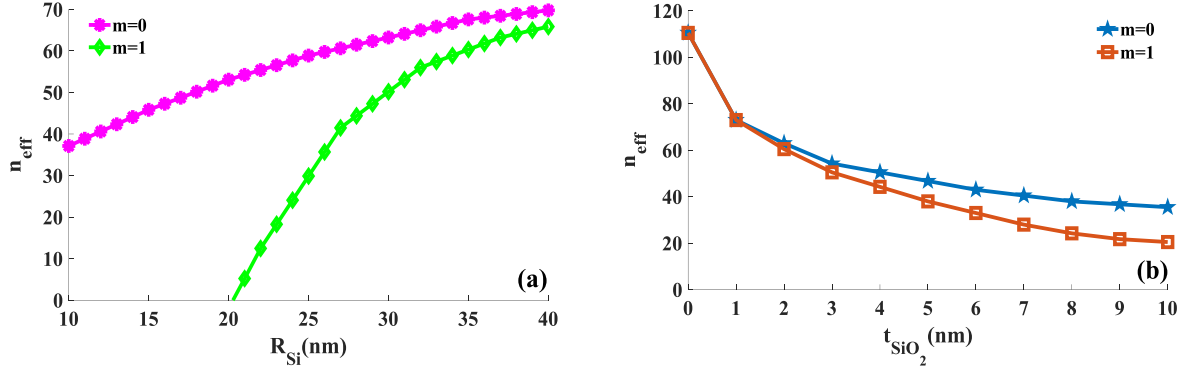

**Fig. 7.** The analytical results of the effective index ($n_{eff}$) as a function of: **(a)** the radius of the Si layer ($R_{Si}$), **(b)** the thickness of the SiO₂ layer ($t_{SiO_2} = R_{SiO_2} - R_{Si}$). The thickness of the SiO₂ in the diagram (a) and the radius of the Si layer in diagram (b) are 3 and 30 nm, respectively ($t_{SiO_2} = 3\ nm, R_{Si} = 30\ nm$). In both diagrams, the chemical potential is 0.7 eV ($\mu_c = 0.7\ eV$), the thickness of the InSb layer is 5 nm ($t_{InSb} = R_{InSb} - R_{SiO_2} = 5\ nm$) and the magnetic bias is supposed to 1 T ($B_0 = 1$). The operation frequency is 35 THz.

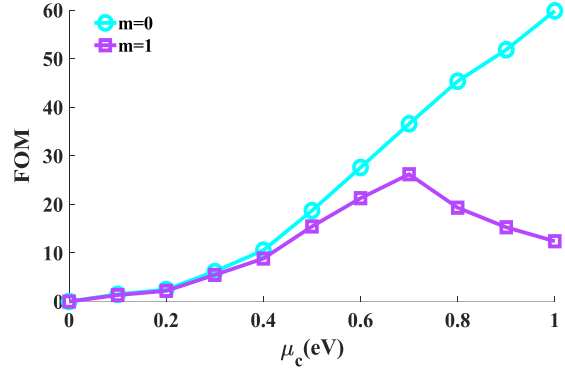

**Fig. 8.** The analytical results of FOM as a function of the chemical potential ($\mu_c$) at the frequency of 35 THz. The magnetic bias is supposed to 1 T ($B_0 = 1$) and the radii of InSb, SiO₂, and Si layers are 38, 33 and 30 nm, respectively ($R_{InSb} = 38\ nm, R_{SiO_2} = 33\ nm, R_{Si} = 30\ nm$ ).

To investigate the dependence of the effective index on the radii of SiO₂ and Si layers, the analytical results of the effective index have been shown as the functions of the SiO₂ radius and Si thickness in Fig. 7. As seen in Fig. 7(a), there is a cut-off radius for the mode $m = 1$, which allows someone to design a single-mode waveguide for $R_{Si} < 20\ nm$. As the radius increases, especially for $R_{Si} > 35\ nm$, the effective index diagrams of two modes become closer. Fig. 7(b) indicates the effective index of two modes as a function of silica thickness. It is clearly observed that the high effective index is achievable for $t_{SiO_2} \to 0$. However, it must be noted that the propagation length decreases for $t_{SiO_2} \to 0$ and thus the mode has very low FOM in this situation. As a result, there is always a trade-off between the effective index and the propagation length for choosing the better silica thickness at the specific frequency.

Now, we consider the effect of the chemical potential on the performance of the plasmonic waveguide. Fig. 8 shows the FOM as a function of the chemical potential. It is evident that the FOM for cut-off free mode ($m = 0$) increases as the chemical potential increases. For the mode $m = 1$, it has an optimum value at the chemical potential of 0.7 eV, FOM reaches to 28. Compared to the conventional graphene-based cylindrical waveguides reported in the literature, our novel waveguide supports hybrid plasmons, with adjustable modal properties by varying the chemical doping and magnetic bias. Moreover, it can be utilized for designing non-reciprocal devices such as Faraday rotation-based components in the THz region.

As a final point, we compare the performance of the proposed structures, with and without graphene layers. Tables 1 and 2 show the effect of the graphene layers on the performance of the proposed structures for the first and second modes, respectively. The operating frequency is 35 THz. The chemical potential of graphene layers in all cases is supposed to be 0.7 eV. In the second structure containing graphene layers, the magnetic bias is 1T. All other parameters remain fixed in both waveguides. These tables clearly indicate that in all cases for the first and the second modes, the proposed waveguides containing graphene layers have better performance than the waveguides without graphene layers. For instance, the FOM of cut-free mode ($m = 0$) for the second structure containing the graphene layer is 33, while it reaches 13.5 for this waveguide in the absence of the graphene layer. It can be seen that the second structure has better performance than the first structure. Consider the FOM of cut-free mode for the first and second structures containing graphene layers. The first structure has a FOM of 21.8 while the second one has FOM of 33 for the first mode. Furthermore, the performance of both waveguides in the first mode is much better than the second mode.

**Table 1.**

Comparison of the performance of two proposed structures for the first mode ($m = 0$), with and without the graphene layer, at the frequency of 35 THz.

| The first structure | | | | | | The second structure | | | | | |
|---|---|---|---|---|---|---|---|---|---|---|---|
| with graphene layer ($\mu_c = 0.7\ ev$) | | | without graphene layer | | | with graphene layer ($\mu_c = 0.7\ ev, B_0 = 1\ T$) | | | without graphene layer | | |
| $n_{eff}$ | $L_{prop}(\mu m)$ | FOM | $n_{eff}$ | $L_{prop}(\mu m)$ | FOM | $n_{eff}$ | $L_{prop}(\mu m)$ | FOM | $n_{eff}$ | $L_{prop}(\mu m)$ | FOM |
| 26 | 2.1 | 21.8 | 14 | 1.4 | 7.84 | 55 | 10 | 33 | 34.3 | 6.6 | 13.5 |

**Table 2.**

Comparison of the performance of two proposed structures for the second mode ($m = 1$), with and without the graphene layer, at the frequency of 35 THz.

| The first structure | | | | | | The second structure | | | | | |
|---|---|---|---|---|---|---|---|---|---|---|---|
| with graphene layer ($\mu_c = 0.7\ ev$) | | | without graphene layer | | | with graphene layer ($\mu_c = 0.7\ ev, B_0 = 1\ T$) | | | without graphene layer | | |
| $n_{eff}$ | $L_{prop}(\mu m)$ | FOM | $n_{eff}$ | $L_{prop}(\mu m)$ | FOM | $n_{eff}$ | $L_{prop}(\mu m)$ | FOM | $n_{eff}$ | $L_{prop}(\mu m)$ | FOM |
| 21 | 1.9 | 15.9 | 11 | 0.9 | 3.96 | 54.1 | 7.27 | 24.3 | 27.6 | 3.2 | 5.2 |

## 4. Conclusion

In this article, a general analytical model has been proposed for anisotropic multi-layer cylindrical structures containing graphene layers. This structure supports tunable plasmons, with adjustable modal features by varying the chemical potential and the magnetic bias. As special cases of the general structure, two exemplary waveguides have been studied, both numerically and analytically. Our investigation is just for the first two modes ($m = 0,1$) to be brief. The first, familiar, structure is composed of the graphene layer deposited on the SiO$_2$ nano-wire. An excellent agreement has been seen between the theoretical and simulation results. In the second example, a novel graphene-based structure with the gyro-electric substrate, constituting Air-Graphene-InSb-SiO$_2$-Si layers, is proposed and investigated. The high field confinement, e.g. the effective index of 100 for $B = 2\ T$ at the frequency of 38 THz, has been obtained for the second structure. The hybridization of the graphene and the gyro-electric substrate in this structure results in tunable non-reciprocal plasmonic features, which is helpful for designing new THz components.

A large value of FOM, amounting to 60 e.g., at the chemical potential 0.95 eV for the mode $m = 0$ is obtained. A comprehensive comparison is done in this paper to investigate the effect of graphene layers on the performance of proposed waveguides. Our results show that the effective index and FOM have larger values for the waveguides incorporating graphene layers compared to the waveguides without graphene layers. Moreover, this comparison indicates that the second structure has larger values of FOM in comparison to the first one, which means that utilizing the gyroelectric substrate together with the graphene layer obtains higher values of FOM. Our presented model of the general structure allows one to design new tunable THz devices, such as modulators, absorbers, and cloaks, by controlling the plasmonic features via the chemical potential and the magnetic bias.